\documentclass[showpacs,floatfix,twocolumn,prl,amsmath,amssymb]{revtex4-1}
\usepackage{amssymb}
\usepackage{epsfig}
\usepackage{graphicx}
\usepackage{dcolumn}
\usepackage{bm}
\usepackage[bookmarks=false]{hyperref}

\begin{document}

\title{Functional renormalization group approach to the dynamics of
first-order phase transitions}
\author{Yantao Li, Fan Zhong}
\email{stszf@mail.sysu.edu.cn}
\affiliation{State Key Laboratory of Optoelectronic Materials and Technologies, School of
Physics and Engineering, Zhongshan University, Guangzhou 510275, People's
Republic of China}
\date{\today}

\begin{abstract}
We apply the functional renormalization group theory to the dynamics of
first-order phase transitions and show that a potential with all odd-order terms can describe spinodal decomposition phenomena. We derive a
momentum-dependent dynamic flow equation which is decoupled from the
static flow equation. We find the
expected instability fixed points; and their associated exponents agree remarkably
with the existent theoretical and numerical results. The complex
renormalization group flows are found and their properties are shown. Both
the exponents and the complex flows show that the spinodal decomposition
possesses singularity with consequent scaling and universality.
\end{abstract}

\maketitle

First-order phase transitions (FOPTs) are ubiquitous in our
diverse physics world. Basically, their dynamics is described by two distinct
mechanisms \cite{Gibbs}: one is nucleation and growth and the other is spinodal
decomposition. FOPTs show discontinuity at their transition points at which the first (and higher) order derivatives of the
chemical potential of the two phases involved change discontinuously. This is in sharp contrast to continuous phase transitions that have singularity at their transition points \cite{fisher1967} and are generally described by the elegant renormalization-group (RG) theory \cite{wilson1974}, Consequently, there is a challenging question of whether FOPTs can also show
any singularity and consequent scaling and universality. Recently, it has been found \cite{zhong2005, zhong2011} that the spinodal
decomposition is possibly controlled by instability fixed points (IFPs) whose
exponents equal those of the Yang--Lee edge singularity \cite{yang1952}. Further evidence for the existence of the IFPs was studied in the
two-dimensional Potts model \cite{Fan2011}. However, more studies
on these new and imaginary fixed points are still needed in particular to search support from other theories and methods.

A possible way to study the existence of the IFPs is the functional
RG (FRG) theory, which, like the perturbative
RG theory, also originates from Wilson's idea of RG \cite{wilson1974} but provides us both perturbative and nonperturbative
aspects. The FRG theory provides a unified picture to describe a lot of phenomena in many different fields \cite{ERG}.
It has also been applied to understand nucleation in the scheme of
Langer's theory \cite{langer}. In these studies \cite{Tetradis}, for a final
potential $U_{k_{f}}\approx m_{k_{f}}^{2}\phi ^{2}/2+\gamma _{k_{f}}\phi
^{3}/6+\lambda _{k_{f}}\phi ^{4}/8$, which is identical with its initial form,
where $m_{k_{f}}$ is the mass and $\gamma _{k_{f}}$ and $\lambda _{k_{f}}$ are
the couplings in the final scale $k_{f}$, a ratio $R(h)$ with $h=9\lambda
_{k_{f}}m_{k_{f}}^{2}/\gamma _{k_{f}}^{2}$ is introduced to determine
the region of validity of the homogeneous nucleation theory. When $\lambda
_{k_{f}}/m_{k_{f}}$ is fixed, the spinodal line is approached from $%
h\rightarrow 0$. This corresponds to the region in which $%
m_{k_{f}} $ and $\lambda _{k_{f}}$ tend to zero simultaneously and hence the
potential is left only with the term $\gamma _{k_{f}}\phi ^{3}/6$. Although the
nucleation theory described by the FRG theory does not valid in this region,
it still indicates the spinodal decomposition has a potential of $\phi ^{3}$ like
with a massless mode, which is just the viewpoint in \cite{zhong2005}. In the FRG perspective, we thus expect that the potential of the spinodal
decomposition contains all odd order terms of the field $\phi $ for
symmetry reason, in comparison with the critical phenomena which contain all even-order terms with $Z_{2}$ symmetry. To confirm this argument, we need to find the expected IFPs in the FRG scheme, determine the corresponding exponents, and then compare them with
extant results.

In the FRG scheme, one needs to solve the exact static Wetterich equation \cite{wetterich1993} with appropriate approximations to determine a fixed point and its
exponents. We shall use the Blaizot, M\'{e}ndez-Galain, and Wschebor
(BMW) approximation \cite{bmw}, which contains full momentum dependence and has yielded excellent results for static critical exponents among others \cite{bmws}. As the spinodal decomposition is kinetic in origin, we have
to consider dynamics. However, dynamic flow equations have only been derived with the derivative expansion method \cite{canet2007}. So, we shall first derive in the BMW scheme a dynamic flow equation as well as the decoupled static BMW equation from a dynamic Wetterich equation.

Consider the Hamiltonian in the bare scale $\Lambda $,%
\begin{equation}
\mathcal{H}=\int_{x}\left\{ \frac{1}{2}\left[ (\nabla \phi)^{2}+m_{\Lambda
}^{2}\phi ^{2}\right] +\frac{1}{3!}\gamma _{\Lambda }\phi ^{3}+\frac{1}{4!}%
\lambda _{\Lambda }\phi ^{4}\right\}.%
\end{equation}%
The dynamic action $\mathcal{S}$ for Model A \cite{hohe} can be expressed in
the form \cite{janssen},
\begin{equation}
\mathcal{S}=\int_{\mathbf{x}}\left\{ \bar{\phi}\left( \partial \phi
/\partial t+D\delta \mathcal{H}[\phi ]/\delta \phi \right) -D\bar{\phi}%
^{2}\right\} ,
\end{equation}%
with a response field $\bar{\phi}$, where $\mathbf{x}=(t,x)$ and $D$ is a kinetic coefficient. To transform to the FRG scheme, we add to $\mathcal{S}$ a
term
\begin{equation}
\Delta \mathcal{S}_{k}=\frac{1}{2}\int_{\mathbf{x}}\Phi \hat{R}_{k}\Phi
^{\tau},\qquad\hat{R}_{k}=DR_{k}\left(
\begin{array}{cc}
0 & 1 \\
1 & 0%
\end{array}%
\right)%
\end{equation}%
with the static regulator $R_{k}(q)=Z_{k}(k^{2}-q^{2})\Theta (k^{2}-q^{2})$ \cite{litim}, where $k$ is the flow scale, $\Phi =(\phi ,\bar{\phi})$, $Z_{k}$ is the renormalization factor for the field $\phi $, $\Theta$ the Heaviside step function, and $\tau$ denotes a transposition.
$\Delta \mathcal{S}_{k}$ plays a role of mass that suppresses
fluctuations of momentum $|q| \lesssim k$. As a result,
the scale-dependent effective action,
\begin{equation}
\Gamma _{k}[\varphi ,\bar{\varphi}]=-W_{k}[h,\bar{h}]-\frac{1}{2}\int_{%
\mathbf{x}}\psi \hat{R}_{k}\psi ^{\tau}+\int_{\mathbf{x}}\sum_{i=1}^{2}h_{i}\varphi _{i},\label{sdea}
\end{equation}%
with $W_{k}[h,\bar{h}]=\ln \int \mathcal{D}\Phi \exp ( -\mathcal{S-}\Delta
\mathcal{S}_{k}+\int_{\mathbf{x}}\sum_{i=1}^{2}h_{i}\phi _{i})$, interpolates between the mean field action $\mathcal{S}[\psi ]=\Gamma_{k=\Lambda}$ and the effective action $\Gamma \lbrack \psi ]=\Gamma _{k=0}$ as fluctuations are progressively taken into account as $k$ flows from the bare microscopic scale $\Lambda$ to the long-distance scale 0, where $\psi =(\varphi ,\bar{\varphi})$ with $\varphi_{i}$ denoting the average of $\phi _{i}$ and $(h,\bar{h})$ is the conjugate source. We have introduced a subscript $i\in\{ 1,2\}$ such that $(\phi _{1},\phi
_{2})=(\phi,\bar{\phi})$ for convenience. Defining the $2\times2$ matrices of the vertex functions as \cite{canet2007}
\begin{eqnarray}
\hat{\Gamma}_{k}^{\left(2\right)}(\mathbf{x}_{1},\mathbf{x}_{2}) &=&
\frac{\delta ^{2}\Gamma _{k}}{\delta \varphi _{i_{1}}(\mathbf{x}_{1})\delta
\varphi _{i_{2}}(\mathbf{x}_{2})},  \label{gamma2}\\
\hat{\Gamma}_{i_{3},\cdots ,i_{n}}^{(n>2)}(\mathbf{x}_{1}\mathbf{,}\cdots ,%
\mathbf{x}_{n}) &=&\frac{\delta ^{\left( n-2\right)}\hat{\Gamma}_{k}^{(2)}(%
\mathbf{x}_{1},\mathbf{x}_{2})}{\delta \varphi _{i_{3}}(\mathbf{x}%
_{3})\cdots \delta \varphi _{i_{n}}(\mathbf{x}_{n})},\label{gamma3}
\end{eqnarray}
one writes the flow of the effective average action as
\begin{equation}
\partial _{s}\Gamma _{k}=\frac{1}{2}\mathrm{Tr}\int_{\mathbf{q}}
\hat{G}_{k}\partial _{s}\hat{R}_{k},%
\label{wetter}
\end{equation}%
in the Fourier space, where $\hat{G}_{k}=[\hat{\Gamma}_{k}^{\left(2\right)}+\hat{R}_{k}]^{-1}$ is the full field-dependent propagator, $\mathbf{q}=(\omega ,q)$, and $s=k\partial _{k}$ is the RG time. Equation~(\ref{wetter}) is just the dynamic Wetterich equation \cite{ERG} reflecting the RG flow of $\Gamma_k$ under an infinitesimal change of $k$. From Eq.~(\ref{wetter}) and using Eq.~(\ref{gamma3}), one can derive the flow equation of two-point vertex function in uniform field configurations \cite{canet2010}
\begin{eqnarray}
\partial _{s}\Gamma _{i,j}^{\left(2\right)}(\mathbf{p})=\mathrm{Tr}
\int_{\mathbf{q}}(\partial _{s}\hat{R}_{k})\hat{G}_{k}(\mathbf{q})\left[-\frac{1}{2}\hat{\Gamma}_{i,j}^{\left(4\right)}(\mathbf{p},-
\mathbf{p},\mathbf{q})\right. \quad\nonumber\\
\left.+\hat{\Gamma}_{i}^{\left(3\right)}(\mathbf{p},\mathbf{q})\hat{G}
_{k}(\mathbf{p}+\mathbf{q})\hat{\Gamma}_{j}^{\left(3\right)}(-\mathbf{p},
\mathbf{p}+\mathbf{q})\right]\hat{G}_{k}(\mathbf{q}),\quad
\label{flow}
\end{eqnarray}%
where $\Gamma _{i,j}^{\left(2\right)}$ are the
matrix elements of $\hat{\Gamma}_{k}^{\left(2\right)}$. In Eqs.~(\ref{gamma2}) to (\ref{flow}), all the $\varphi_i$ dependence are implicit.

In order to derive from Eq.~(\ref{flow}) the specific dynamic and static flow equations for Model A, one needs considering only those vertex functions that comply with its structure. Since only $\Gamma _{\Lambda}^{\left( 1,1\right)}=i\omega +D(q^{2}+m_{\Lambda}^{2})$, $\Gamma
_{\Lambda}^{\left(2,1\right)}=D\gamma _{\Lambda }$, $\Gamma _{\Lambda}^{\left( 3,1\right)}=D\lambda _{\Lambda }$, and $\Gamma _{\Lambda}^{\left(0,2\right)}=-2D$ exist at $\varphi _{i}=0$ in the bare scale, we set other vertex functions zero \cite{note1} in Eq.~(\ref{flow}). To decouple the flow equations for $\partial _{s}\Gamma _{2,2}^{(2)}(\mathbf{p})$ and the real part of $\partial
_{s}\Gamma _{1,2}^{(2)}(\mathbf{p})$, we drop the $\omega$ dependence in $\Gamma _{k}^{\left(0,2\right)}$ and assume
\begin{equation}
\Gamma_{k}^{\left(1,1\right)}\left(\mathbf{q}\right) =i\frac{\omega}{2D}%
\Gamma_{k}^{\left(0,2\right)}\left(q\right) +D\Gamma _{k}^{\left(2\right)}\left(q\right),%
\label{ward}
\end{equation}%
where $\Gamma _{k}^{(n)}(q)$ is the static vertex
function. Equation~(\ref{ward}) is a kind of Ward identity expressing the fluctuation-dissipation
theorem. It becomes exact when the $\omega $ dependence of $\Gamma _{k}^{(0,2)}(
q) $ is re-included and no mode
couplings are present \cite{folk2006}. By Eq.~(\ref{ward}), vertex functions of higher orders such as $\Gamma _{k}^{(2,1)}$ and $\Gamma _{k}^{(3,1)}$ depend no longer on $\omega $ either. In addition, the external frequency is set to zero, since we mainly focus on $\mathbf{p}=(0,p)$. After
performing the trace and integrating over $\omega $, one finds the static
Wetterich equation about $\Gamma _{k}^{(2)}(p) $
decoupled from the dynamic behavior. Then, the approximation
$\Gamma _{k}^{(n+1)}( q_{1},\cdots ,q_{n},0) =$
$\partial _{\varphi }\Gamma _{k}^{(n)}( q_{1},\cdots
,q_{n}) $ results in the BMW equation \cite{bmw}%
\begin{eqnarray}
\partial_{s}\Gamma_{k}^{(2)}(p)
=\int_{q}\left[\partial _{s}R_{k}(q)\right] G_{k}^{2}(q) \quad\qquad\qquad\qquad\qquad\qquad\nonumber\\
\times\left\{G_{k}(p+q)\left[\partial_{\varphi}\Gamma
_{k}^{(2)}(p)\right]^{2}-\frac{1}{2}\partial _{\varphi}^{2}\Gamma
_{k}^{(2)}(p)\right\},\qquad%
\label{all}
\end{eqnarray}%
and a momentum-dependent dynamic flow equation
\begin{eqnarray}
\partial _{s}\Gamma _{k}^{(0,2)}(p)
=-\int_{q}[\partial _{s}R_{k}( q)] G_{k}^{2}(q)G_{k}(p+q)\Gamma
_{k}^{(0,2)}(q) \nonumber\\
\times \Gamma _{k}^{(0,2)}(p+q)
\left[\partial _{\varphi}\Gamma _{k}^{(2)}( p)
\right]^{2}(A+2B)(A+B)^{-2},\qquad%
\label{dyflow}
\end{eqnarray}%
which is one of our main results, where $G_{k}(q)=[\Gamma _{k}^{(2)}(q) +R_{k}(
q)]^{-1}$, $A\equiv \Gamma _{k}^{(0,2)}( q)
G_{k}^{-1}(p+q)$, and $B\equiv \Gamma _{k}^{(0,2)}( p+q)
G_{k}^{-1}(q)$. Note that the dynamics part does not contribute to Eq.~(\ref{all}) and the fixed points are thus solely determined by the latter as should be.

Having these flow equations, one needs to drop their
dimensions to access the fixed points. The dimensionless renormalized variables and
functions are defined as
\begin{equation}
\begin{array}{l}
\tilde{p}=p/k,\qquad \tilde{q}=q/k,\qquad \tilde{\varphi}%
=[k^{2-d}Z_{k}K_{d}^{-1}]^{1/2}\varphi,  \\
\tilde{U}_{k}=k^{d}U_{k},\text{ \ \ }\tilde{\bar{\varphi}}=[k^{-2-d}%
\bar{Z}_{k}]^{1/2}\bar{\varphi},\text{ \ \ }\tilde{D}=Z_{D}D, \\
\tilde{\Gamma}_{k}^{\left( 2\right)}=[k^{2}Z_{k}]^{-1}\Gamma _{k}^{\left(
2\right)},\text{ \ }\tilde{\Gamma}_{k}^{\left(0,2\right)}=-[2Z_{D}\bar{Z}%
_{k}D]^{-1}\Gamma _{k}^{\left(0,2\right)},%
\end{array}
\label{dimensionless}
\end{equation}%
where $U_{k}$ is the potential, $K_{d}^{-1}\equiv d 2^{d-1}\pi^{d/2}\Gamma( d/2)$ ($\Gamma$ is the Euler Gamma
function), and $Z_{D}$ and $\bar{Z}_{k}$ are the renormalization factor of $D$
and $\bar{\varphi}$, respectively, and are related through $Z_{k}^{1/2}=Z_{D}\bar{Z}_{k}^{1/2}$. In the scaling regime, $Z_{D}\sim k^{-\varepsilon _{k}}$ and $Z_{k}\sim k^{-\eta _{k}}$, with $\varepsilon _{k}$ and $\eta _{k}$ being related to the dynamic and the static critical exponents by $z=2-\varepsilon _{k=0}$ and $\eta =\eta _{k=0}$, respectively. Consequently, the dimensionless dynamic flow equation satisfies
\begin{equation}
\partial _{s}\tilde{\Gamma}_{k}^{\left(0,2\right)}(\tilde{p})=\left( \eta
_{k}-\varepsilon _{k}+\tilde{p}\partial _{\tilde{p}}\right)  \tilde{\Gamma}_{k}^{\left(0,2\right)}(\tilde{p})+\partial
_{s}\Gamma _{k}^{\left(0,2\right)}(p),
\label{ddflow}
\end{equation}%
where the last term is Eq.~(\ref{dyflow}) in its dimensionless form. The static flow equation can be separated into
two dimensionless flow equations: One is the flow equation of $\tilde{U}_{k}^{\prime \prime}=\tilde{\Gamma}_{k}^{\left(2\right)}(
\tilde{0})$ (a prime denotes a derivative with $\tilde{\varphi}$) obtained by setting $p=0$ in Eq. (\ref{all}), the other is the
flow equation of $\chi _{k}=Z_{k}(\varphi )/Z_{k}(0)$ which comes
from $\partial _{s}Z_{k}=[\partial (\partial _{s}\Gamma _{k}^{\left(2\right)}(p))/\partial p^{2}]_{p=0}$. They are \cite{guerra}
\begin{eqnarray}
\partial _{s}\tilde{U}_{k}^{\prime\prime}&=&(\eta _{k}-2)\tilde{U}%
_{k}^{\prime\prime}+\frac{1}{2}(d-2+\eta _{k})\tilde{\varphi}\tilde{U}%
_{k}^{\prime\prime\prime} \nonumber\\
&&-\left(1-\frac{\eta _{k}}{d+2}\right)\left[
\tilde{G}_{k}^{2}\tilde{U}_{k}^{\prime\prime\prime\prime}-2\tilde{G}%
_{k}^{3}(\tilde{U}_{k}^{\prime\prime\prime})^{2}\right],%
\label{uflow}\\
\partial _{s}\chi _{k}&=&\eta _{k}\chi _{k}+\frac{1}{2}(d-2+\eta _{k})\tilde{%
\varphi}\chi _{k}^{\prime}-(\tilde{U}_{k}^{\prime\prime\prime})^{2}%
\tilde{G}_{k}^{4} \nonumber\\
&&+\left(1-\frac{\eta _{k}}{d+2}\right)\left(4\tilde{G}_{k}^{3}\tilde{U}%
_{k}^{\prime\prime\prime}\chi _{k}^{\prime}-
\tilde{G}_{k}^{2}\chi _{k}^{\prime\prime}\right)
\label{zflow}
\end{eqnarray}
for our choice of $R_k$, where $\tilde{G}_{k}=(1+\tilde{U}_{k}^{\prime\prime})^{-1}$.

From the dimensionless dynamic and static flow equations~(\ref{ddflow}) to (\ref{zflow}), one can find the IFPs and their
exponents in a way similar to the critical phenomena. The only difference is that we need to change to the symmetry that describes the spinodal decomposition. We use the field
expansion method to find the IFPs \cite{guerra,canet2003}. The key point is how to retain terms in $U_{k}$ and $Z_{k}$. As mentioned above, the potential should be of the form
\begin{equation}
\tilde{U}_{k}=\lambda _{2}\tilde{\varphi}^{2}+\lambda _{3}\tilde{\varphi}%
^{3}+\lambda _{5}\tilde{\varphi}^{5}+\cdots +\lambda _{n_{1}}\tilde{\varphi}%
^{n_{1}}+\cdots.%
\label{newu}
\end{equation}%
The expansion of $Z_{k}$ (or $\chi _{k}$) is delicate. Any of its $\tilde{\varphi}$-dependent term  will bring even besides odd order terms to the \emph{unrenormalized} potential. In this potential, however, the even-order terms that arise from the even-order terms of $Z_{k}$ are generated by the coupling of the latter terms with $\lambda_{2}$ only, which vanishes at the mean-field or bare-scale spinodal point. Consequently, in order to have a potential of odd-order couplings at the bare scale, we retain only the even-order terms in $Z_{k}$ and write
\begin{equation}
\chi _{k}=1+\chi _{2}\tilde{\varphi}^{2}+\chi _{4}\tilde{\varphi}%
^{4}+\cdots +\chi _{n_{2}}\tilde{\varphi}^{n_{2}}+\cdots.%
\label{newz}
\end{equation}%
Substituting Eqs.~(\ref{newu}) and (\ref{newz}) into Eqs.~(\ref{uflow}) and (%
\ref{zflow}), one gets a set of nonlinear flow equations for the coefficients $\lambda _{n_{1}}$ and $\chi _{n_{2}}$, which reach their fixed points at $\partial _{s}\lambda _{n_{1}}=\partial _{s}\chi _{n_{2}}=0$. To obtain the exponents, we use $\tilde{p}=0$ and $\tilde{\varphi}=0$ as the renormalization
point and let $\tilde{\Gamma}_{k}^{(0,2)}( \tilde{q}) =1$ in Eq.~(\ref{dyflow}). With this approximation, Eq.~(\ref{ddflow}) recovers the dynamic flow equation in Ref.~\cite{canet2007} if the $\tilde{\varphi}$ dependence of the latter equation is neglected, which confirms ours. From Eqs.~(\ref{zflow}) and (\ref{ddflow}), one finds
\begin{eqnarray}
\eta _{k}&=&\left.\frac{ \tilde{G}_{k}^{2}\chi _{k}^{\prime\prime}-4\tilde{G}_{k}^{3}\tilde{U}_{k}^{\prime\prime\prime}\chi _{k}^{\prime}+(\tilde{U}_{k}^{\prime\prime\prime})^{2}\tilde{G}_{k}^{4}}
{ \chi _{k}+\frac{1}{d+2}\tilde{G}_{k}^{2}\chi
_{k}^{\prime\prime}-\frac{4}{d+2}\tilde{G}_{k}^{3} \tilde{U}_{k}^{\prime\prime\prime} \chi _{k}^{\prime}}\right|_{\tilde{\varphi}=0},\label{eta}\\
z_{k}&=&\left. 2-\eta _{k}+\frac{3}{2}\left(1-\frac{\eta _{k}}{d+2}\right)(\tilde{U}
_{k}^{\prime\prime\prime})^{2}\tilde{G}_{k}^{4}\right\vert _{\tilde{\varphi}=0},
\label{z}
\end{eqnarray}
respectively. In order to compare with direct numerical results, we also compute the hysteresis exponents~\cite{zhong2005,zhong2011},
\begin{equation}
n_{H}=\frac{d+2-\eta}{d+2-\eta+2z},\quad n_{m}=\frac{d-2+\eta}{d+2-\eta
+2z},  \label{hyster}
\end{equation}%
which reflect the hysteresis of the coercivity and the remnant magnetization,
respectively.

\begin{figure}[tp]
 \centerline{\epsfig{file= 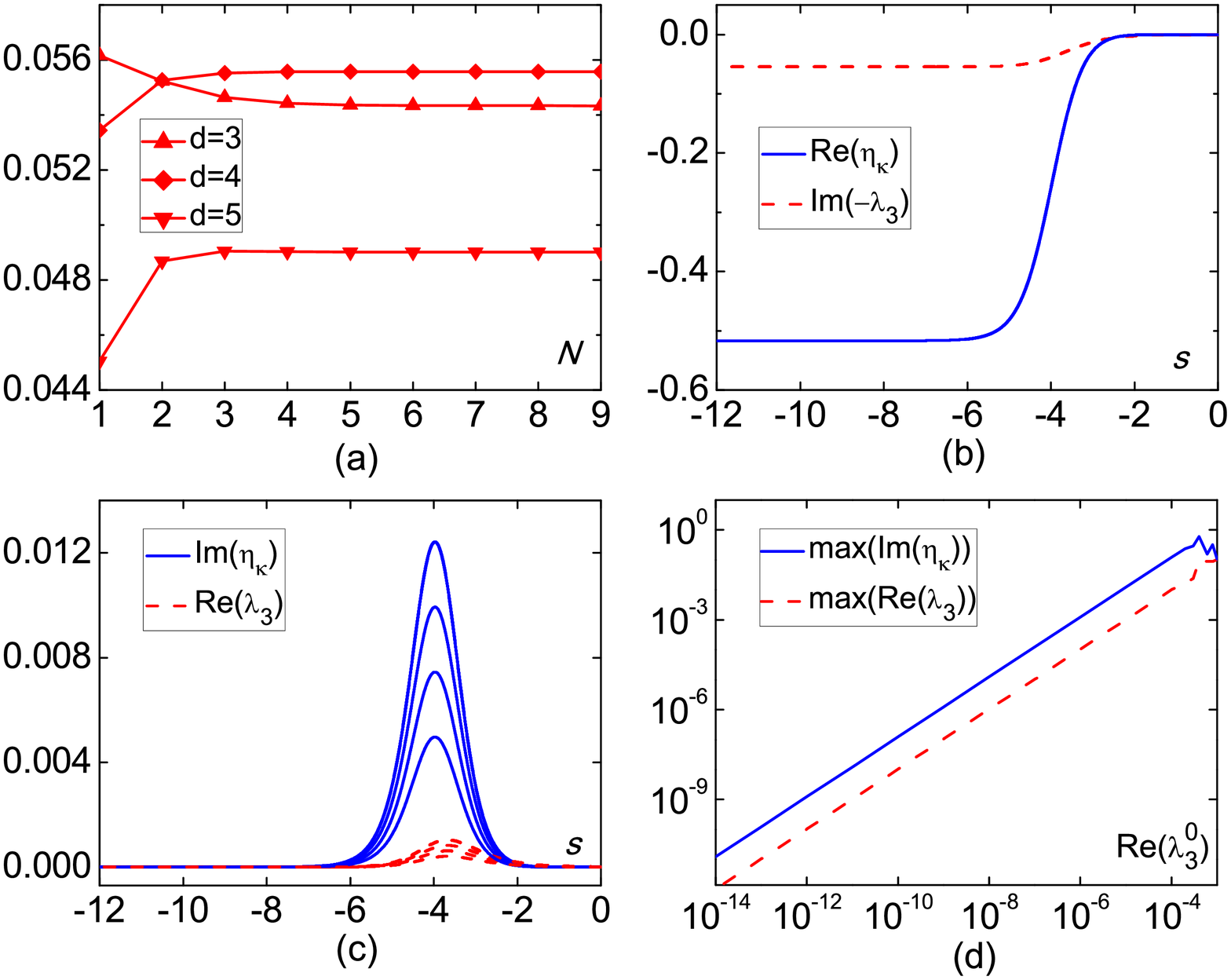,height=6.6cm,width=9.6cm}}
 \caption{\label{final} (Color online) (a) Stability of Im$(\protect\lambda %
_{3})$ with increasing $N$ in $d=3,4,5$. RG flows of (b) Re$(\protect\eta _{k})$ and Im$(\protect\lambda _{3})$ and (c)
Im$(\protect\eta _{k})$ and Re$(\protect\lambda _{3})$ from $s=0
$ to $s=-12$ with four different values of Re$(\protect\lambda %
_{3}^{0})$: $0.1\times 10^{-4}$, $0.8\times 10^{-5}$, $0.6\times 10^{-5}$, $%
0.4\times 10^{-5}$ (top to bottom). Whereas the flows in (b) do not depend on
Re$(\protect\lambda _{3}^{0})$, those in (c) are proportional to Re$
(\protect\lambda _{3}^{0})$ and can thus overlap when scaled by the proportionality. (d) The solid line and the dashed line
are the maximum of Im$(\protect\eta _{k})$ and Re$(\protect\lambda _{3})$, respectively. Both of them are linear with Re$(\protect\lambda _{3}^{0})$. This relation is broken with larger Re$(\protect\lambda _{3}^{0})$. In (b), (c), and (d), $d=3$ and $N=9$.}
\end{figure}
We now present the results for the fixed points. For $d\geq6$, the only infrared-stable fixed point found is a Gaussian fixed point with $\lambda _{n_1}=0$, confirming the mean-field behavior in these dimensions. For $d<6$ on the other hand, a lot of new fixed points appear. To find the nontrivial IFPs, we start with $n_{1}=3$ and $n_{2}=0$. There are only three fixed points: Besides the Gaussian one, the others are just the IFPs of a pair of two purely imaginary conjugate values of $\lambda _{3}$. This can be checked by their stability against increasing $n_{1}$ and $n_{2}$. When $n_{1}$ is fixed, the IFPs become stable with increasing $n_{2}$, whose maximum value is $n_{1}-3$ for closing the set of equations. So, the stability can be checked by $n_{1}=2N+1=3,\ 5,\cdots $ and $n_{2}=n_{1}-3$. One sees from Fig.~\ref{final}(a) that the imaginary part of $\lambda_{3}$, denoted by Im$(\lambda _{3})$, indeed stabilizes quickly as $N$ increases.

\begin{table}[tp]
\caption{Instability Exponents: $^{a}$ This work. $^{b}$ Perturbative RG
Results \protect\cite{borel, breuer}. $^{c}$ Numerical results \protect\cite%
{zhong2005}.}
\label{table}
\begin{ruledtabular}
\begin{tabular}{ccccc}
$d$ & $6$ & $5$ & $4$ & $3$ \\
\hline
$\eta ^{a}$ & $0$ & $-0.147$ & $-0.328$ & $-0.517$ \\
$\eta ^{b}$ & $0$ & $-0.147$ & $-0.328$ & $-0.524$ \\
$z^{a}$ & $2$ & $1.922$ & $1.809$ & $1.661$ \\
$z^{b}$ & $2$ & $1.938$ & $1.874$ & $1.809$ \\
$n_{H}^{a}$ & $2/3$ & $0.650$ & $0.636$ & $0.624$ \\
$n_{H}^{c}$ & $0.655(2)$ & $0.645(3)$ & $0.625(12)$ & $0.595(30)$ \\
$n_{m}^{a}$ & $1/3$ & $0.260$ & $0.168$ & $0.0546$ \\
$n_{m}^{c}$ & $0.33(1)$ & $0.27(4)$ & $0.17(8)$ &  \\ 
\end{tabular}%
\end{ruledtabular}
\end{table}
Having determined the IFPs, we can then extract their exponents. We note first that if we neglect the dependence of $\chi_k$ on $\tilde{\varphi}$ from the beginning, we find $\eta=-0.176$, $-0.427$, and $-0.669$ for $d=5$, $4$, and $3$, respectively, which are already quite good compared with those obtained from a perturbative RG theory \cite{borel} shown in Table~\ref{table}. Taking into account the field expansion of $\chi_k$ spoils, however, the results. Yet, if we set $\chi _{2}=0$ in Eqs.~(\ref{eta}) and (\ref{z}), the resultant exponents given in Table \ref{table} agree remarkably with the extant ones. Two remarks are in order here. First, the perturbative results were derived only to two- and three-loop orders and thus may not be accurate. Second, our $z$s are a bit smaller than the others, this may arise either from our setting $\tilde{\Gamma}
_{k}^{\left(0,2\right)}(\tilde{q})=1$ in Eq.~(\ref{dyflow}), or our omitting the $\tilde{\varphi}$ dependence in Eq.~(\ref{ddflow}). Still, the remarkable agreement in all the three dimensions considered indicates that the condition of $\chi_2=0$ is not accident. A consequence is that it leads to a relation between $\lambda _{2}$ and $\lambda _{3}$ via $\eta_k$ from Eqs.~(\ref{uflow}) and (\ref{zflow}), the relation which may be pertinent to the fact that the $\varphi^{3}$ theory has only one independent static exponent. Although further studies are clearly needed, the good agreements still confirm our identification of the imaginary fixed points with the IFPs.

To show how to reach the imaginary IFPs from the bare scale, we fix the value of $\lambda _{2}$ to its fixed point value at any scale, and set Im$(\lambda _{n_{1}\geq 3}^{0})=10^{-16}$, Re$(\lambda _{n_{1}>3}^{0})=0$, and $\chi _{n_{2}}^{0}=0$ at the bare scale indicated by the superscripts, while varying Re$(\lambda_{3}^{0})$. Then, we
solve the set of nonlinear flow equations of $\lambda _{n_{1}}$ and $\chi
_{n_{2}}$ up to $N=9$ in $d=3$ to $5$. The flows start from the bare scale $%
\Lambda $ $(s=0)$ to the zero scale $(s\rightarrow -\infty )$, see Fig.~\ref%
{final}(b) and (c). The flows of Re$(\eta _{k})$ and Im$%
(\lambda _{3})$ do not depend on Re$(\lambda _{3}^{0})$; they overlap
completely for different Re$(\lambda _{3}^{0})$s. The flows of Im$(\eta _{k})$ and Re$(\lambda _{3})$, on the other hand, depend on Re$(\lambda
_{3}^{0})$. Nevertheless, they are proportional to
Re$(\lambda _{3}^{0})$. In other words, if Re$(\lambda
_{3}^{0}) $ is multiplied by a constant, then the flows can still overlap after
multiplying all values of the flows by the same constant. If the
value of Re$(\lambda _{3}^{0})$ is too large, however, the proportional
relation is broken [see Fig.~\ref{final}(d)], and the IFP becomes
unreachable. Although this appears to indicate that the IFP has a controlling region, it becomes larger as Im$(\lambda _{n_{1}}^{0})$ increases. In addition, flows of purely real $\lambda _{n_{1}}^{0}$s cannot reach the fixed points in our numerical analysis, though those from analytical solutions may. Therefore, we see that the IFPs are indeed infrared stable and reachable. A tiny initial complex coupling appears sufficient for the flows to reach the fixed point in the FRG theory. This confirms a previous perturbative analysis \cite{zhong2011}.
As the complex flows connecting the physics at the bare scale with the physics at the
zero scale are a process of adding fluctuations to the mean-field theory, which is controlled by the Gaussian fixed point and describes the spinodal decomposition well for $d\geq6$ \cite{zhong2005,zhong2011}, the FRG theory thus shows that the same phenomena in $d<6$ should be
controlled by the IFP, though the initial purely imaginary coupling may have to be identified.

In summary, we have applied the FRG theory to the dynamics of FOPTs. The momentum-dependent dynamic flow equation
in the BMW scheme has been derived. We have found the IFPs with their exponents in good agreement with extant results albeit with some subtleties. This gives a physical meaning of spinodal decomposition for a potential
with all the odd-order terms as compared with the critical phenomena for a potential with all the even-order terms. The complex flows and their properties have also been shown in the FRG scheme. Both the exponents and the complex flows show that the spinodal decomposition has a behavior of
singularity and consequent scaling and universality.

This work was supported by the NSF of PRC (No.10625420).

\end{document}